# Limits of isotropic plastic deformation of Bangkok clay

P. Evesque
Lab MSSMat,  UMR 8579 CNRS, Ecole Centrale Paris
92295 CHATENAY-MALABRY, France, e-mail: evesque@mssmat.ecp.fr

**Abstract:**
*A model assuming incremental plastic isotropic response has been recently proposed to model the deformation of isotropic packing of grains, in the small-strain range. It is used here on overconsolidated remould clay, to interpret the small-strain range behaviour obtained in [1,2] on Bangkok clay. The data published in [1,2] at constant volume are also used here to measure the size of the domain of validity in the $(q/(M'p), p/p_o)$ plane, where $p_o$ is the over-consolidation isotropic pressure, $p=(\sigma_1+\sigma_2+\sigma_3)/3$ is the mean stress and q the deviatoric stress, $q=\sigma_1-\sigma_3$. So, it is shown that the model works also for clay. This enlarges the application domain of model [3,4] to soft clay with OCR larger than 1.2 to 1.5.*

Recently, I have proposed to describe the early stage of stress–strain $\sigma$-$\varepsilon$ behaviour of an initially isotropic packing of grains using an incremental plastic modelling with isotropic response hypothesis [3, 4]. According to it, one can write:

$$\begin{pmatrix} \delta\varepsilon_1 \\ \delta\varepsilon_2 \\ \delta\varepsilon_3 \end{pmatrix} = \frac{1}{C_o} \begin{pmatrix} 1 & -\nu & -\nu \\ -\nu & 1 & -\nu \\ -\nu & -\nu & 1 \end{pmatrix} \begin{pmatrix} \delta\sigma_1 \\ \delta\sigma_2 \\ \delta\sigma_3 \end{pmatrix} \quad (1a) \qquad \text{with} \qquad \nu = 1 - \sigma_1/\{2(1+M)\sigma_2\} \quad (1b)$$

In the above equations, the parameter $\nu$ has been adjusted so that the deformation law obeys the Rowe's dilatancy relation when the compression is made at $\sigma_2=\sigma_3$=constant; hence M is the friction coefficient written in terms of the ratio between the deviatoric stress $q=\sigma_1-\sigma_3$ and lateral stress $\sigma_3$, i.e. $M= q/\sigma_3$ at critical state; so $M= 2\sin(\varphi)/[1-\sin(\varphi)]$, where $\varphi$ is the friction angle at critical state. Eq. 1b could have been written in terms of the friction ratio $M'=q/p$ at critical state; in this case, $p=(\sigma_1+\sigma_2=\sigma_3)/3$ is the mean stress, and one gets $M=3M'/(3-M')$ or $M'=3M/(3+M)$, with $M'=6\sin(\varphi)/[3-\sin(\varphi)]$. In Eq. 1, $\nu$ and $1/C_o$ depend on the stress; they are not Poisson's coefficient nor Young's modulus, since the deformation is plastic and irreversible instead of elastic; hence we call them pseudo Poisson's coefficient and pseudo Young's modulus.

This is indeed the simplest modelling the small-strain range; it only requires isotropic response; it shall work at the beginning of deformation, till the deformation generates anisotropy.

At larger deformation one expects however that the classical Coulomb criteria is obeyed in the case of granular materials and soils, so that the flow rule is no more isotropic at this stage, but obeys the one of a theory of plasticity, with a single mechanism involved.

As it is quite general, the same modelling, i.e. Eq. 1a, shall work also for any kind of material as far as its response is isotropic; of course, Eq. 1b may not hold always, since it expresses the Rowe's law that is specific to granular matter and clay. But if it is not satisfied, stability condition of the pile under isotropic compression requires $\nu < 1/2$ t at q=0.

On the other hand, in many studies on clay, the camclay model is often proposed as the reference model to interpret experimental data on clays at the early stage of deformation; I think it is a pity owing to the following reasons:

## Some criticisms about using camclay modelling of clay behaviour:

(i) in the large strain limit, it turns out that the camclay modelling fails as soon as the axi-symmetry of the experiment is not achieved, which is encountered most of the time: basically, camclay way of thinking imposes some importance to symmetry; so, it incites writing the yielding rules in a way similar to the Drücker-Pragger law when the two minor or major stresses $\sigma_2, \sigma_3$ are different ($\sigma_2 \neq \sigma_3$). On the contrary, real experimental flow rules obey the real Coulomb's formulation rather well, that shows up a symmetry breaking between the roles plaid by $\sigma_2$ and $\sigma_3$ and that gives little importance to the intermediate principal stress, and leading importance to maximum and minimum principal stresses.

For instance, criteria that writes equivalent role plaid by $\sigma_2$ and $\sigma_3$ lead to completely incompatible prediction of the limit of slope stability, as shown in [5], (indeed, stability of slopes inclined at 50° are computed in [5] using Dr¨cker-pragger law with materials having a critical friction angle equal to 30° and with no help of dilatancy effect).

(ii) In the domain of small deformation, the isotropic nature of the response which is currently observed at this stage, is attributed to the elastic behaviour of the material most of the time. But one knows undoubtedly that irreversible deformation are produced in this range of strain, and its amplitude is of the order of the whole strain.

(iii) In the domain of the intermediate strain range, one is forced to use more complex flow rule with more parameters, that ensures a better modelling. A typical examples is the Pender's model [6], which has been used to interpret the mechanical behaviour of Bangkok clay studied in [1,2]. Such a modelling is complex in general and requires computer assisted prediction.

So, according to these inadequacies, camclay can not represent correctly stress-strain behaviour in the domains of both small and large deformations except in the region of normally consolidated samples, for which it has been conceived; it is then a pity that it is still so often used, and that it serves as a "benchmark" model, from which many new modelling start. Indeed, one sees quite often in the literature that the small strain domain of isotropic response is "adequately" fitted using the elastic modelling entering camclay, while the true experimental behaviour is known not to be elastic because strain is irreversible.

By contrast the simplicity of Eq. 1, which just implies the isotropy of the response, may be useful to predict main trends at the early stage of the deformation when the sample is isotropic initially. As told earlier, it applies not only to granular matter but also to any kind of isotropic material, in particular on remould clays. Hence it is useful and important to test and find the domain of validity of this modelling. Some examples have been given and discussed in [7-9] for some other clay. Here I study the case of the Bangkok clay, whose behaviour has been reported in the literature [1,2] and interpreted in terms of Pender's model [6].

A simple way to determine the domain of validity of the isotropic modelling is to perform an undrained compression, which imposes no volume change, i.e. $\delta v=0$, as it is shown below:

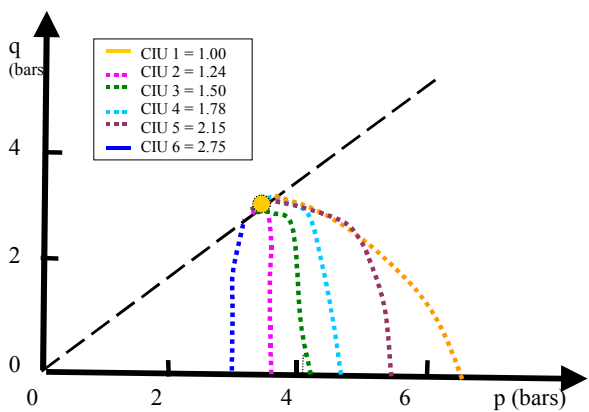

**Figure 1:** *Typical plots of trajectories of undrained tests for over-consolidated isotropic samples of Bangkok clays* ciu=$p_i/p_o$, *as deduced from ref* [1], *in the* (q,p) *plane.* q=$\sigma_1-\sigma_3$; 3p=$\sigma_1+\sigma_2+\sigma_3$. *Pressure of overconsolidation* $p_o$; $p_o$ *range: 6.5-8 bars* [1]; *initial pressure of the undrained tes are indicated in the figure using* ciu*:* $p_i$= $p_o$/ciu, *with* ciu=1, 1.24, 1.5, 1.78, 2.15 & 2.75. *The inclined line corresponds to the friction law* q=M'p.

Fig. 1 reports typical trends of undrained triaxial compression tests performed on isotropic samples of Bangkok clay at different values of the over consolidation pressure, as can be found in [1]. It is recalled that ciu is the ratio between the over-consolidation pressure $p_o$ and the initial pressure $p_i$ at which the triaxial compression starts with undrained condition in the present case. Under such a test, $\delta v=0$ ; so, the isotropic modelling of Eq. 1 imposes that the mean pressure p remains also constant or that $v= ½$ . Eq. (1b) tells that $v= ½$ occurs when $\sigma_1-\sigma_2=M$, i.e. when the stress is on the characteristic line q=M'p.

If Eq. (1b) does not hold, sample stability of an isotropic material tells that $v< ½$ , so that isotropic assumption imposes that the trajectory of an undrained test is vertical in the (q,p) plane.

When the material obeys Eq. (1b), $v$ evolves with stress; the complete prediction with isotropic hypothesis is the following: the trajectory remains vertical in the (q,p) plane, i.e. p=constant, till it meets the characteristic line q=M'p ; then the trajectory turns right or left and follows the characteristic line q=M'p [3,4]; the trajectory ends at the critical point. The characteristic line is the dashed line in Fig. 1, and the critical point is the yellow point. However it is difficult to conclude to the validity of the isotropic modelling when the path follows the characteristic lines, because it is not the unique possibility and because it is surely wrong at the end of the test, since the material obeys the Coulomb criterion at this stage, that is not isotropic; but one notes that trajectories of material with ciu>2.5 follows exactly the predicted trends.

Anyway, for all these reasons, it is safer to consider that isotropic modelling remains valid when the trajectory in the (q,p) plane remains vertical. Hence, Fig. 1 shows that isotropic plastic modelling of Bangkok clay is valid for $q<Mp_{ci}$ as soon as ciu>1.3 ; $p_{ci}$ is the ending pressure of the undrained test when ciu<1.8, or it is the pressure at which the trajectory hits the characteristic line for ciu>2. Also, one can note that the last part of the trajectory in the (q,p) plane is rather horizontal when 1.3<ciu<1.8; this is incompatible with the isotropic modelling; but it is also not well fitted by the modelling used in [1]. At last, camclay modelling is undoubtedly better when ciu<1.3 in the case of undrained test.

As a conclusion, one notes that isotropic elastic modelling is used in [1] to interpret the domain of small-q results; but it is known that the true response of a clay is irreversible, so that the present modelling which uses Eq. 1 is better. One notes also that the irreversible modelling used in [1] , i.e. the one which describes the "non isotropic response", does not fit efficiently the experimental data in the (q,p) plane, while it seems to work rather correctly in the $(q,\varepsilon_1)$ plane; but a correct modelling requires a good fitting along both planes, so that it can not be viewed as satisfactory.

The model proposed by Eq. 1 works efficiently in the case of undrained test; it has been also checked on oedometric path, i.e. $\varepsilon_2=\varepsilon_3=0$. It can be important to test it on other paths imposing some other relation between $\varepsilon_1$ and $\varepsilon_2=\varepsilon_3$., i.e. $\varepsilon_2=k\ \varepsilon_1$ .

*Acknowledgements:* CNES is thanked for partial funding.

The electronic arXiv.org version of this paper has been settled during a stay at the Kavli Institute of Theoretical Physics of the University of California at Santa Barbara (KITP-UCSB), in june 2005, supported in part by the National Science Fundation under Grant n° PHY99-07949.

*Poudres & Grains* can be found at :
http://www.mssmat.ecp.fr/rubrique.php3?id_rubrique=402